\newcommand{\beq}{\begin{equation}}
\newcommand{\eeq}{\end{equation}}
\newcommand{\sh}{{\rm sh}}
\newcommand{\ch}{{\rm ch}}
\newcommand{\id}
 {i\kern.06em\hbox{\raise.25ex\hbox{$/$}\kern-.60em$\partial$}}

\newcommand{\bs}{/\kern-.52em b}
\newcommand{\qs}{/\kern-.52em s}

\newcommand{\p}{\partial}

\newcommand{\dd}
{\kern.06em\hbox{\raise.25ex\hbox{$/$}\kern-.60em$\partial$}}

\documentstyle[12pt]{article}
\date{}
\begin{document}
\title{Gaussian Effective Potential Analysis of Sinh(Sine)-Gordon
 Models by New Regularization-Renormalization Scheme
\footnotetext{\# Corresponding author}
\footnotetext{* On leave of absence from the Physics Department,
Shanghai University, Shanghai, 201800}}
\author{{Sze-Shiang Feng$^{1,2,\#,*}$, and Guang-Jiong Ni$^{3}$}\\
1. {\small {\it CCAST (World Lab.),P.O. Box. 8730, 100080, Beijing}}\\
2.  {\small {\it Modern Physics Department, University of Science
           and Technology of China, 230026, Hefei, China,}}\\
       e-mail:sshfeng@yahoo.com\\
3.   {\small {\it Physics Department, Fudan University
    200433, Shanghai, China}}\\e-mail:gjni@fudan.ac.cn\\}
\maketitle
\newfont{\Bbb}{msbm10 scaled\magstephalf}
\newfont{\frak}{eufm10 scaled\magstephalf}
\newfont{\sfr}{eufm7 scaled\magstephalf}
\input amssym.def
\baselineskip 0.3in
\begin{center}
\begin{minipage}{135mm}
\vskip 0.3in
\baselineskip 0.3in
\begin{center}{\bf Abstract}\end{center}
  {Using the new regularization and renormalization scheme recently
  proposed by Yang and used by Ni et al , we analyse the
  sine-Gordon and sinh-Gordon
  models  within the framework of Gaussian effective potential in
  $D+1$ dimensions. Our
  analysis suffers no divergence and so does not suffer from the
  manipulational obscurities in the conventional analysis of divergent
  integrals. Our main conclusions agree exactly with
  those of Ingermanson
  for $D=1,2$ but disagree for $D=3$: the $D=3$ sinh(sine)-Gordon
  model
  is non-trivial. Furthermore, our analysis shows that for $D=1,2$, the
  running coupling constant (RCC)has poles for sine-Gordon
  model($\gamma^2<0$) and
  the sinh-Gordon
  model ($\gamma^2>0$) has a possible critical point $\gamma^2_c$
  while for $D=3$,
  the RCC has poles for both $\gamma^2>0$ and $\gamma^2<0$.
  \\PACS number(s): 11.10.-z,11.10.Gh.
   }
\end{minipage}
\end{center}
\vskip 1in
\begin{center}
\section{Introduction}
\end{center}
The "Gaussian effective potential" (GEP) has proven to be a powerful
non-perturbative approach in quantum field theories (QFT). Using the
GEP approach, Stevenson etc. found two distinct, non-trivial versions
of the 3+1 dimensional $\lambda\phi^4$ theory: the "precarious
$\phi^4$ theory" and the "autonomous $\phi^4$ theory"\cite{s1}, and thus
provided a new view point about the triviality
 of $\lambda\phi^4$ model as a  physical theory. Also
by GEP, Ingermanson examined the generalized sinh-Gordon and sine-Gordon
model in $D+1$ dimensions\cite{s2}. The Lagrangian for the model takes
in general the form
\beq
{\cal L}=\frac{1}{2}(\p_{\mu}\phi)^2-\frac{m^2}{\gamma^2}[{\rm ch}
(\gamma\phi)-1]
\eeq
where $m$ and $\gamma$ are the  mass and  coupling constant
respectively at tree level. If $\gamma^2>0$,the classical potential
is a cosh curve
with a single minimum at the origin; if $\gamma^2<0$, it is actually
a sine-Gordon model with an infinite number of degenerate minimum
of the potential. The limiting case $\gamma^2\rightarrow 0$ is
usually understood
to be a free theory of masss $m$. When $D=1$, the sine-Gordon model
is equivalent to a group of other models\cite{s3}, namely,  the
massive
Thirring model\cite{s4}, the Coulomb gas\cite{s5},  the continuum
limit of the $xyz$ spin$=\frac{1}{2}$ model\cite{s6} and
the massive O(2) non-linear $\sigma$-model\cite{s5}. It is convenient
to define $\beta^2=-\gamma^2$ for discussing the sine-Gordon model.
It has been shown that the $D=1$ sine-Gordon model is
superrenormalizable
for $0\leq\beta^2\leq 4\pi$; renormalizable for $4\pi\leq
\beta^2\leq
8\pi$,and nonrenormalizable for $\beta^2>8\pi$\cite{s7}, the last
property was first discovered by Coleman\cite{s4}. Based on GEP,
Ingermanson
concluded that for $D\geq 3$, the model (1) can exist only as a free
theory while for $D<3$, the vacuum is unstable over a certain range
of the coupling constant. In Ingermanson's analysis, the integrals
\beq
I^D_n(\mu^2)=\int\frac{d^D{\bf p}}{(2\pi)^D}\frac{\sqrt{{\bf p}^2
+\mu^2}}{({\bf p}^2+\mu^2)^n}
\eeq
may be divergent or finite. The divergent ones were dealt with without
using any cutoff procedure or regularization procedure  and were just
taken
to be as though finite most of the time, and the whole
analysis seems to be  regularization scheme
independent. Yet for $D\geq 3$, the fact that
$I^D_2(\mu^2)$ is divergent was used to lead to the conclusion that
the interacting theory is inconsistent for $D\geq 3$. Hence, the rule
that taking $I^D_n$ as finite was violated here and there exists
such a kind of manipulational obscurity.\\
\indent To eliminate this obscurity, we intend to re-analyse the model
(1) by the new regularization and renormalization (R-R) scheme,
which was proposed by
Yang\cite{s8} and used by Ni et al recently\cite{s9}-\cite{s12}
though the "derivative regularization" trick has been evolving in
the literatures for many years
\cite{s13}-\cite{s18}.
 The spirit is like this:
when encountering a superficially
divergent Feynmann diagram integral (FDI), we first differentiate
it with
respect to some parameter such as a mass parameter enough times until
it becomes convergent and the integration can be done. Then we
reintegrate
it with respect to the same parameter the same times. The result
is to be
taken as the definition of the original FDI. Then instead of
divergence,
some arbitrary constants appear in FDI.
The appearence of these arbitrary constants
indicates some lack of
theoretical knowledge
about the model at QFT level under consideration. The determination
of them is beyond the ability of the QFT, instead, they should
be fixed by experiment via some suitable renormalization
procedure. This new R-R scheme has turned out to be successful in
that
the whole analysis is quite clearcut and it can give a prediction
of Higgs mass, $m_H=138$ GeV in the standard model\cite{s11}.
Also it provides an elegant calculation in QED, e.g. Lamb shift
\cite{s12}
. In this paper our main conclusions agree exactly with those of
Ingermanson
  for $D=1,2$. But for $D=3$ there is an important discrepancy
  : the $D=3$ sinh(sine)-Gordon model
  may be non-trivial. Furthermore, our analysis shows that for
  $D=1,2$, the
  running coupling constant (RCC)has poles for $\gamma^2<0$ and
  the sinh-Gordon
  model has a possible critical point $\gamma^2_c$ while for $D=3$,
  the RCC has poles for both $\gamma^2>0$ and $\gamma^2<0$.
 In section 2, we give a general analysis
of the model (1) in the Schr$\ddot o$dinger representation and
present some
known results. In section 3, we analyse the model for $D=1,2,3$
respectively
by the new R-R scheme. The last section is devoted to discussions.
\begin{center}
\section{General Analysis}
\end{center}
\subsection{GEP and running coupling constant(RCC)}
The Lagrangian (1) can be rewritten as
\beq
{\cal L}=\frac{1}{2}\dot{\phi}^2-V(\phi)
\eeq
\beq
V(\phi)=\frac{1}{2}(\p_i\phi)^2+\frac{m^2}{\gamma^2}({\rm ch}
\gamma\phi-1)
\eeq
The canonical momentum conjugate to $\phi$ is
\beq
\pi=\frac{\p{\cal L}}{\p\dot{\phi}}=\dot{\phi}
\eeq
and the Hamiltonian reads
\beq
H=\int d^D{\bf x}[\frac{1}{2}\pi^2+V(\phi)]
\eeq
The quantization is realized through
\beq
[\pi(x_0,{\bf x}), \phi(x_0, {\bf y})]=-i\delta^D({\bf x}-{\bf y})
\eeq
which can be satisfied if
\beq
\pi=-i\frac{\delta}{\delta\phi}+G(\phi)
\eeq
In particular, we often choose $G(\phi)=0$. In Schr$\ddot o$dinger
representation, the state is described by wave functional $\Psi[\phi]$
which satisfies the Schr$\ddot o$dinger equation
\beq
H\Psi[\phi]=E\Psi[\phi]
\eeq
The first step in Gaussian variational method is to make an ansatz
for the Schr$\ddot o$dinger wave functional for the vacuum
\beq
\Psi[\phi, \Phi, {\cal P},f]={\cal N}_f\exp
\{i\int{\cal P}_{\bf x}\phi_{\bf x}-\frac{1}{2}\int_{{\bf x,y}}
(\phi_{\bf x}-\Phi_{\bf x})f_{{\bf x,y}}(\phi_{\bf y}-\Phi_{\bf y})\}
\eeq
The ${\cal P},\Phi,f$ are variational parameters.
 The energy of the
variational state eq.(10) is
\beq
E[\Phi, {\cal P}, f]\equiv <\Psi\mid H\mid\Psi >
=\int_{\bf x}\{\frac{1}{2}{\cal P}^2_{\bf x}+\frac{1}{2}
(\p_i\Phi)^2+\frac{m^2}{\gamma^2}[Z_{\bf x}{\rm ch}\gamma\Phi-1]
+\frac{1}{4}[f_{{\bf xx}}-\int_{\bf y}\delta_{{\bf xy}}
\nabla_{\bf x}^2 f^{-1}_{{\bf xy}}]\}
\eeq
where
\beq
Z_{\bf x}\equiv \exp(\frac{1}{4}\gamma^2 f^{-1}_{{\bf xx}})
\eeq
We are interested in finding the effective potential, so we consider
the energy of the state with constant classical field $\Phi, \p_i\Phi=0$.
The extremum energy configuration clearly satisfies the constraint
${\cal P}=0$. The variational equation
\beq
\frac{\delta E}{\delta f_{{\bf xy}}}=0
\eeq
gives the general forms of $f_{{\bf xy}}$ and $f_{{\bf xy}}^{-1}$ as
\beq
f_{{\bf xy}}=\int \frac{d^D{\bf p}}{(2\pi)^D}\sqrt{{\bf p}^2+\mu^2}\cos
{\bf p}\cdot({\bf x}-{\bf y})
\eeq
\beq
f^{-1}_{{\bf xy}}=\int \frac{d^D{\bf p}}{(2\pi)^D}
\frac{\cos{\bf p}\cdot({\bf x}-{\bf y})}{\sqrt{{\bf p}^2+\mu^2}}
\eeq
Using $I^D_n(\mu^2)$ in eq.(2), we have (we often omit the superscript
$D$)\beq
f_{{\bf xx}}=I_0(\mu^2),\enspace\enspace\,\,\,\,\,
f_{{\bf xx}}^{-1}=I_1(\mu^2)=2\frac{\p}{\p\mu^2}I_0
\eeq
The energy density ${\cal E}$ is a function of $\Phi$ and $\mu^2$
\beq
{\cal E}(\Phi,\mu^2)=\frac{1}{2}I_0(\mu^2)-\frac{1}{4}\mu^2I_1(\mu^2)
+\frac{m^2}{\gamma^2}[Z_m(\mu^2){\rm ch}\gamma\Phi-1]
\eeq
where
\beq
Z_m(\mu^2)=\exp[\frac{1}{4}\gamma^2I_1(\mu^2)]
\eeq
According to Ritz variational principle\cite{s19}, any stationary state
(10) is an eigenstate of the discrete spectrum of $H$, and the
corresponding
eigenvalue is the stationary value of the function (17). Thus
we consider the stationary points $(\bar{\mu}^2,\bar{\Phi})$ for
${\cal E}$ which are solutions of the equations
\beq
\frac{\p{\cal E}}{\p\Phi}=\frac{m^2}{\gamma}Z_m(\mu^2){\rm sh}\gamma
\Phi=0
\eeq
\beq
\frac{\p{\cal E}}{\p\mu^2}=\frac{1}{8}I_2(\mu^2)[\mu^2-m^2Z_m(\mu^2)
{\rm ch}
\gamma\Phi]=0
\eeq
i.e
\beq
\sh\gamma\bar{\Phi}=0
\eeq
\beq
\bar{\mu}^2=m^2Z_m(\bar{\mu}^2)\ch\gamma\bar{\Phi}
\eeq
(As one is interested in the effective potential, one may consider
the
stationary point $\bar{\mu}^2$ and leave $\Phi$ free as we will do
in the
following.) Clearly, if $\gamma^2>0$, $\bar{\mu}^2$ is always
positive and we have the only solution $(\bar{\mu}^2,\Phi=0)$. Instead, if
$\gamma^2=-\beta^2<0$, $\bar{\mu}^2$ is positive only when
$\cos\beta\bar{\Phi}>0$, so it is necessary that
$(2n-\frac{1}{2})\pi\leq\beta\bar{\Phi}\leq(2n+\frac{1}{2})\pi,
(n\in\Bbb N)$.
but eq.(21) confines it to be $\sin\beta\bar{\Phi}=0$. So we have
an infinite
number of stationary points $(\bar{\mu}^2, \bar{\Phi}_n=2n\pi$). It
is evident that for all stationary points, the energy takes the same
value. Therefore, for negative $\gamma^2$, the stationary states
are infinitely degenerate.\\
\indent To guarantee that the stationary point is an local minimum,
we have
to demand that the matrix
\beq
{\cal M}=\left(\matrix{\frac{\p^2{\cal E}}{\p\Phi^2}&\frac
{\p^2{\cal E}}{\p\Phi
\p(\mu^2)}\cr
\frac{\p^2{\cal E}}{\p(\mu^2)\p\Phi}&\frac{\p^2{\cal E}}{\p(\mu^2)^2}
\cr}\right)
\eeq
is positively definite. Since
\beq
\frac{\p^2{\cal E}}{\p\Phi\p(\mu^2)}=-\frac{1}{8}\gamma m^2I_2
Z_m(\mu^2)\sh\gamma\Phi
\eeq
\beq
\frac{\p^2{\cal E}}{\p(\mu^2)^2}=\frac{1}{8}I_2-\frac{3}{16}
\mu^2I_3+\frac{3}{16}m^2I_3Z_m(\mu^2)\ch\gamma\Phi
+\frac{1}{64}\gamma^2m^2I_2^2Z_m(\mu^2)\ch\gamma\Phi
\eeq
\beq
\frac{\p^2{\cal E}}{\p\Phi^2}=m^2Z_m(\mu^2)\ch\gamma\Phi
\eeq
we have from (21) and (22)
\beq
\frac{\p^2{\cal E}}{\p(\mu^2)^2}_{\mid \bar{\mu}^2,\bar{\Phi}}=
\frac{1}{8}I_2(1+\frac{1}{8}\bar{\mu}^2\gamma^2I_2)
\,\,\,\,\,\,\,\,\,\,\,
\frac{\p^2{\cal E}}{\p\phi^2}_{\mid\bar{\mu}^2,\bar{\Phi}}=\bar{\mu}^2,
\,\,\,\,\,\,\,\,\,\,\,
\frac{\p^2{\cal E}}{\p\Phi\p(\mu^2)}_{\mid \bar{\Phi},\bar{\mu}^2}=0
\eeq
So for ${\cal M}$ to be positively definite, we should have
\beq
\frac{1}{8}I_2(1+\frac{1}{8}\bar{\mu}^2\gamma^2I_2)>0
\eeq
\indent The GEP is defined as
\beq
V_G(\Phi)={\cal E}(\Phi,\mu^2(\Phi))
\eeq
where the fuctional relation of $\mu^2$ to $\Phi$ is the same as (22)
of $\bar{\mu}^2$ to $\bar{\Phi}$. Like the usual effective
potential $V_{eff}$ obtained by loop expansions\cite{s20},
$V_G$ has also the physical interpretation: it is the minimum
of the expectation value of the energy density for all
states constrained by the condition
that the field $\phi$ has expectation value $\Phi$.
Using (22), $V_G$ can be written as
\beq
V_G=\frac{1}{2}I_0(\mu^2)-\frac{1}{4}\mu^2I_1(\mu^2)
+\frac{\mu^2-m^2}{\gamma^2}
\eeq
It is straight forward to check that
\beq
\frac{dV_G}{d(\mu^2)}=\frac{1}{\gamma^2}(1+\frac{1}{8}\mu^2\gamma^2
I_2(\mu^2))
\eeq
\beq
\frac{d\mu^2}{d\Phi}=\gamma\mu^2{\rm th}\gamma\Phi[1+\frac{1}{8}
\mu^2\gamma^2I_2(\mu^2)]^{-1}
\eeq
and so
\beq
\frac{dV_G}{d\Phi}=\frac{\mu^2}{\gamma}{\rm th}\gamma\Phi
\eeq
Clearly, $V_G$ acquires its minimum at $\Phi_0=0$, which agrees with
$\bar{\Phi}$. (In general, the stationary points of an arbitrary
function $f(x,y)$ agree with those of $f(x(y),y)$, where
$x$ as a function of $y$ is determined by $\p f/\p x=0$, but whether
$f(x,y)$ and $f(x(y),y)$ acquire their maximum or minimum
simultaneously just depends.)\\
\indent  For later use, we calculate the following derivatives. First
we have
\beq
\frac{d^2V_G}{d\Phi^2}=\frac{1}{\gamma}\frac{d\mu^2}{d\Phi}{\rm th}
\gamma\Phi+\frac{\mu^1}{\gamma}\frac{d{\rm th}\gamma\Phi}{d\Phi}
\eeq
From (32) and $d{\rm th}\gamma\Phi/d\Phi=\gamma/\ch^2\gamma\Phi$, we
have

\newcommand{\F}{1+\frac{1}{8}\mu^2\gamma^2I_2}
\newcommand{\th}{{\rm th}}
\newcommand{\I}{\frac{1}{8}\gamma^2I_2-\frac{3}{16}\mu^2\gamma^2I_3}

\beq
\frac{d^2V_G}{d\Phi^2}=\mu^2{\rm th}^2\gamma\Phi(\F)^{-1}+
\mu^2\ch^{-2}\gamma\Phi
\eeq
$$
\frac{d^3V_G}{d\Phi^3}=\gamma\mu^2\th^3\gamma\Phi(\F)^{-2}+3\gamma
\mu^2
\frac{\th\gamma\Phi}{\ch^2\gamma\Phi}(\F)^{-1}-$$
\beq
\gamma(\mu^2)^2\th
^3\gamma\Phi(\F)^{-3}(\I)-2\gamma\mu^2\ch^{-3}\gamma\Phi\sh\gamma\Phi
\eeq
$$
\frac{d^4V_G}{d\Phi^4}=\gamma^2\mu^2\th^4\gamma\Phi(\F)^{-3}
+6\gamma^2\mu^2\frac{\th^2\gamma\Phi}{\ch^2\gamma\Phi}(\F)^{-2}$$
$$-4(\gamma\mu^2)^2\th^4\gamma\Phi(\F)^{-4}(\I)
+3\gamma^2\mu^2\frac{1-2\sh^2\gamma\Phi}{\ch^4\gamma\Phi}(\F)^{-1}$$
$$-6(\gamma\mu^2)^2\frac{\th^2\gamma\Phi}{\ch^2\gamma\Phi}(\F)^{-3}
(\I)$$
$$+3\gamma^2(\mu^2)^3\th^4\gamma\Phi(\F)^{-5}(\I)^2
$$
$$-\gamma^2(\mu^2)\th^4\gamma\Phi(\F)^{-4}(-\frac{3}{8}\gamma^2I_3
+\frac{5}{32}\mu^2\gamma^2I_4)$$
\beq
-2\gamma^2\mu^2\frac{\th^2\gamma\Phi}{\ch^2\gamma\Phi}(\F)^{-1}-
2\gamma^2\mu^2(\ch^{-2}\gamma\Phi-\frac{3\th^2\gamma\Phi}{\ch^2\gamma
\Phi})
\eeq
At $\Phi_0, \ch\gamma\Phi_0=0$, we have
\beq
\frac{d^4V_G}{d\Phi^4}_{\mid \Phi_0}=3\gamma^2\mu^2(\F)^{-1}
-2\gamma^2\mu^2
\eeq
\indent The renormalization is carried out at $\Phi_0$ (it will be
referred to as $\Phi_0$-renormalization) and the renormalized
mass and coupling constant
are defined by
\beq
m^2_R\equiv\frac{d^2V_G}{d\Phi^2}_{\mid\Phi_0}
\eeq
\beq
m^2_{R}\gamma^2_{R}\equiv\frac{d^4V_G}{d\Phi^4}_{\mid\Phi_0}
\eeq
We see from (40) that the renormalization of the coupling constant
depends
on that of the mass. We deduce from (35) and (38) that
\beq
m^2_R=\mu^2
\eeq
\beq
m^2_R\gamma^2_R=3\gamma^2\mu^2(\F)^{-1}-2\gamma^2\mu^2
\eeq.
Eq(41) just asserts that the renormalized mass, which is in general
the energy difference of one-particle state and the vacuum
\cite{s21},
equals the variational parameter.
\subsection{The Running Coupling Constant}
\indent  Analogous to that in the $\lambda\phi^4$ model\cite{s11},
the running
coupling constant (RCC) is defined
\newcommand{\K}{1+\frac{1}{8}\mu^2\gamma^2I_2\ch^{-2}\gamma\Phi}
$$\gamma^2[\mu^2(\Phi)]=\frac{d^4V_G}{d\Phi^4}/\frac{d^2V_G}{d\Phi^2}
$$
$$=\gamma^2\th^4\gamma\Phi(\F)^{-2}(\K)^{-1}$$
$$+6\gamma^2\th^2\gamma\Phi\ch^{-2}\gamma\Phi
(\F)^{-1}(\K)^{-1}$$
$$-4\gamma^2\mu^2\th^4\gamma\Phi(\F)^{-3}(\K)^{-1}(\I)$$
$$+3\gamma^2(1-2\sh^2\gamma\Phi)\ch^{-4}\gamma\Phi(\K)^{-1}$$
$$-6\gamma^2\mu^2\th^2\gamma\Phi\ch^{-2}\gamma\Phi(\F)^{-2}
(\I)(\K)^{-1}$$
$$+3\gamma^2(\mu^2)^2\th^4\gamma\Phi(\F)^{-4}(\I)^2(\K)^{-1}$$
$$-\gamma^2(\mu^2)^2\th^4\gamma\Phi(\F)^{-3}
(-\frac{3}{8}\gamma^2I_3+\frac{5}{32}\mu^2\gamma^2I_4)(\K)^{-1}$$
$$-2\gamma^2\th^2\gamma\Phi\ch^{-2}\gamma\Phi(\K)^{-1}$$
\beq
-2\gamma^2(1-3\th^2\gamma\Phi)\ch^{-2}\gamma\Phi
(\F)(\K)^{-1}
\eeq
It can be easily seen that $\gamma^2[\mu^2(\Phi)]$ has poles at
\beq
\ch\gamma\Phi=0
\eeq
\beq
\F=0
\eeq
and
\beq
\K=0
\eeq
The poles corresponding to eqs(44)-(46) are of the fourth, first and
the fourth order respectively.
\begin{center}
\section{The New R-R Analysis}
\end{center}
\subsection{The $D=1$ Case}
Following the spirit of the new regularization, we have
\beq
I_2=\frac{1}{\pi\mu^2}\,\,\,\,\,\,\,\,\,
I_1=-\frac{1}{2\pi}\ln\frac{\mu^2}{\mu^2_s}\,\,\,\,\,\,\,\,\,\,\,\,
I_0=C-\frac{\mu^2}{4\pi}(\ln\frac{\mu^2}{\mu_s^2}-1)
\eeq
where $C, \mu_s^2$ are two arbitrary constants. It can be easily
seen that
only $\mu^2_s$ is non-trivial and is to be determined by some
renormalization scheme. Thus we only need the mass renormalization
condition. We choose such a scheme
that {\sl the} $\Phi_0$-{\sl renormalized mass
is just the mass given at the tree level}, i.e.
\beq
m_R^2=m^2
\eeq
So from (22) and (41) we have
$Z_m=1$ which fixes $I_1=0$, thus $\mu_s^2=m^2$. Consequently,
the renormalized coupling constant $\gamma_R^2$ is
\beq
\gamma_R^2=\frac{1-\frac{1}{4\pi}\gamma^2}{1+\frac{1}{8\pi}\gamma^2}
\gamma^2
\eeq
i.e. the coupling constant endures a finite renormalization which can
provide
us with some important information about the model after quantization.
Since it is
usually expected that quantum corrections are small so
$\gamma^2_R$ and $\gamma^2$ should be of the same sign, we should have
that
\beq
\frac{1-\frac{1}{4\pi}\gamma^2}{1+\frac{1}{8\pi}\gamma^2}>0
\eeq
which implies that
\beq
-8\pi<\gamma^2<4\pi
\eeq
On the other hand, the optimal $\bar{\Phi}$ for ${\cal E}(\mu^2,
\Phi)$ incidentally coincides with the minimum $\Phi_0$ for $V_G$,
from (28) we have
\beq
\gamma^2>-8\pi
\eeq
So the two conditions agree well and confirm that there exists a critical
value for $\gamma^2$, i.e. for $\gamma^2>0,\gamma^2<4\pi$,but
for $\gamma^2=-\beta^2<0,\beta_c^2=8\pi$.
It seems that for sinh-Gordon model, $\gamma^2=4\pi$
is also a critical point at which $\gamma^2_R=0$,
but whether the higher vertices also become zero, i.e.
whether the model becomes a free one has to be confirmed
by further analysis.\\
\indent Consider now the low-lying excited states relative to
$\Psi[\mu^2
(\Phi),\Phi]$. The gap equation (22) now turns out to be
\beq
(\frac{\mu^2}{m^2})^{1+\frac{\gamma^2}{8\pi}}=\ch\gamma\Phi
\eeq
Since $I_1=-\frac{1}{2\pi}\ln\frac{\mu^2(\Phi)}{m^2}$, we must
have $\mu^2(\Phi)\geq m^2$ for $\gamma^2>0$ and $\mu^2\leq m^2$ for
$0\leq-\gamma^2<8\pi$. That is the mass parameter at tree level
provides a lower bound
for the particle mass of low-lying excited states if $\gamma^2>0$
while
an upper bound if $\gamma^2<0$ after the model is quantized.\\
\indent Now let us see the RCC. It has poles whose locations
are determined by
\beq
1+\frac{1}{8\pi}\gamma^2=0
\eeq
\beq
1+\frac{1}{8\pi}\gamma^2\ch^{-2}\gamma\Phi=0
\eeq
\beq
\ch\gamma\Phi=0
\eeq
So only when $\gamma^2=-\beta^2<0$ does the RCC possess poles at
\beq
\mu^2_1=0,\,\,\,\,\,\,\,\,\,
\mu^2_2=m^2(\frac{\beta^2}{8\pi})^{\frac{1}{2(1-\frac{\beta^2}
{8\pi})}}
\eeq
As $\beta^2\rightarrow\beta^2_c, \mu_2^2\rightarrow
\frac{1}{\sqrt{e}}m^2$.
Thus we see that there appears another mass scale $\mu^2_2$ in the
model. Since the kinks and anti-kinks have masses
$ M_0\sim (\frac{m^2}{\beta^2})^{1/(2-\beta^2/4\pi)}$ and
the breathers have masses
$M_n=2M_0 {\rm sin}(n\frac{\pi\beta^2}{16\pi-2\beta^2})$ in the
sine-Gordon model\cite{s22}-\cite{s23}, it seems that the
mass scale $\mu^2_2$ has nothing to do with the soliton masses.
\subsection{D=2 case.}
\indent Now the regularized integrals are
$$
I_2=\frac{1}{2\pi}(\mu^2)^{-1/2}$$
\beq
I_1=-\frac{1}{2\pi}(\mu^2)^{1/2}+C_1,\,\,\,\,\,\,\,\,\,\,\,\,\,
I_0=-\frac{1}{6\pi}(\mu^2)^{3/2}+\frac{1}{2}C_1\mu^2+C_0
\eeq
$C_0 $ and $C_1$ are two arbitrary constants and only $C_1$ is
nontrivial
as in the
$D=1$ case. So we need only to fix the mass renormalization condition.
Similarly we
have $I_{1\mid\Phi_0}=0$ and so $C_1=\frac{1}{2\pi}(m^2)^{1/2}$.
Hence
the renormalized coupling constant is
\beq
\gamma^2_R=\frac{1-\frac{\gamma^2}{8\pi}m}{1+
\frac{\gamma^2}{16\pi}m}
\gamma^2
\eeq
Similar to eq.(50) we should have
\beq
\frac{1-\frac{\gamma^2}{8\pi}m}{1+\frac{\gamma^2}{16\pi}m}>0
\eeq
so
\beq
-\frac{16\pi}{m}<\gamma^2<\frac{8\pi}{m}
\eeq
From (28) we also have
\beq
1+\frac{m}{16\pi}\gamma^2>0
\eeq
As in the $D=1$ case, we have a critical value for $\beta^2$,
$\beta^2_c=
\frac{16\pi}{m}$ and $\gamma^2=\frac{8\pi}{m}$ seems also to be
a possible critical
point for sinh-Gordon model. As to the low-lying excited states,
from the gap
equation
\beq
\frac{\mu^2}{m^2}=\exp[\frac{1}{8\pi}\gamma^2(m-\mu)]\ch\gamma\Phi
\eeq
we have for $\gamma^2>0$,
\beq
\frac{\mu^2}{m^2}\geq\exp[\frac{1}{8\pi}(m-\mu)]
\eeq
which means $\mu^2\geq m^2$, whereas for $\gamma^2<0$, we
have $\mu^2\leq m^2$.\\
\indent The RCC has poles determined by the equations
\beq
\ch\gamma\Phi=0
\eeq
\beq
1+\frac{1}{16\pi}\mu\gamma^2 =0
\eeq
\beq
(\frac{\mu^2}{m^2})^2\exp[\frac{1}{4\pi}\gamma^2(\mu-m)]
+\frac{1}{16\pi}
\mu\gamma^2=0
\eeq
(where we take $\mu>0$). So
the poles are $\mu_1^2=0,
\mu_2^2=(\frac{16\pi}{\beta^2})^2$ and $\mu^2_3$, which
is determined
by the last
equation (67). These poles exist only for $\gamma^2<0$. Thus after
quantization we have two
mass scales $\mu^2_2$ and $ \mu^2_3$ apart from the mass parameter $m$
at tree level.
\subsection{D=3 Case}
\indent Now the regularized integrals $I_n$ are
$$ I_3=\frac{1}{6\pi^2\mu^2},\,\,\,\,\,\,\,\,\,\,\,\,\,\,
I_2=-\frac{1}{4\pi^2}\ln\frac{\mu^2}{\mu_s^2}$$
$$ I_1=\frac{1}{8\pi^2}\mu^2(\ln\frac{\mu^2}{\mu^2_s}-1)+C_2$$
\beq
I_0=\frac{1}{32\pi^2}\mu^4(\ln\frac{\mu^2}{\mu^2_s}-\frac{3}{2})
+\frac{1}{2}C_2\mu^2+C_3
\eeq
where $\mu_s^2, C_2$ and $C_3$ are arbitrary constants and  $C_3$ is
trivial
. So we need both mass renormalization and coupling constant
renormalization.
According to the renormalization scheme (48) we have also
$I_{1\mid\mu=m}=0$. So
\beq
C_2=\frac{1}{8\pi^2}m^2(1-\ln\frac{m^2}{\mu_s^2})
\eeq
Therefore from (40) we have
\beq
\gamma^2_R=\frac{1+\frac{1}{16\pi^2}m^2\gamma^2\ln\frac{m^2}{\mu^2_s}}
{1-\frac{1}{32\pi^2}m^2\gamma^2\ln\frac{m^2}{\mu^2_s}}\gamma^2
\eeq
To fix $\mu_s^2$ we choose the same scheme as the mass renormalization
: {\sl the} $\Phi_0$-{\sl  renormalized coupling
constant equals the coupling
constant at tree level}:$\gamma^2_R=\gamma^2$.
So we have $\gamma^2=0$ or $\mu_s^2=m^2$. The
first case  is trivial  and can not
determine $\mu_s^2$ . So only the second is of physical significance.
Thus we arrive at an important conclusion that {\sl the $D=3$ sinh(sine)
-Gordon
model is non-trivial}. This is an important descrepancy between our
analysis and that of Ingermanson.\\
\indent The bounds for the particle mass of the
low-lying excited states can also be obtained. From the gap equation and
the fact that
\beq
I_1(\mu^2)=\frac{1}{8\pi^2}\mu^2\ln\frac{\mu^2}{m^2}+
\frac{1}{8\pi^2}(m^2-\mu^2)
\eeq
we have
\beq
\ln\frac{\mu^2}{m^2}=\frac{\gamma^2\mu^2}{32\pi^2}\ln\frac{\mu^2}{m^2}
+\frac{\gamma^2m^2}{32\pi^2}(1-\frac{\mu^2}{m^2})+\ln\ch\gamma\Phi
\eeq
If we define $x\equiv\mu^2/m^2$ and $\kappa\equiv\gamma^2 m^2
/(32\pi^2)$, then the gap equation (72) can be written as
\beq
\ln x=\kappa (x\ln x+1-x)+\ln {\rm ch}\gamma\Phi
\eeq.
Consider the solution of this equation by graphical means. First
when $\gamma^2>0$, for $\phi=0$, the curve of the l.h.s. will
intersect that of the r.h.s. at two points: $x_1=1$ and a larger
$x_2$. As $\Phi$ increases, the first root increases and the
second one decreases. At some critical $\Phi_{cri}$, the two
will meet. As $\Phi$ increases further, there will be no root for
$0<x<\infty$. For $\Phi<\Phi_{cri}$ , in order
to guarantee the local minimun of
of ${\cal E}$, the root must satisfy eq.(28), i.e.
$\kappa\ln x>1$ and $I_2\not=0$. Therefore, for $\Phi=0$,
$x=1$ is not definitely the local minimum.
In general we have that when
$\mu^2(\Phi)\geq m^2$ .For $\gamma^2<0$, there is only
one root of the gap equation.
In this case, if $\beta\Phi=2n\pi$, the root $x=1$ is not either
the local minimum. Since $\ln{\rm cos}\beta\Phi\leq 0,$
we have $x\leq 1$. Certainly, eq(28) must be also
satisfied at the root if it is a local minimum of
${\cal E}$.\\
\indent The analysis of RCC is a little more difficult. Eq(44) gives a
pole
$\mu_1^2=0$ when $\gamma^2<0$. Eq(45) now reads
\beq
1-\frac{\gamma^2m^2}{32\pi^2}\frac{\mu^2}{m^2}\ln\frac{\mu^2}{m^2}=0
\eeq
Since for $\gamma^2>0, \mu^2\geq m^2$, there is only one solution to it.
For $0<-\gamma^2<\frac{32\pi^2e}{m^2}, \mu^2\leq m^2$, there will exist
two solutions. Eq(46) can also give one  pole for the $\gamma^2>0$ and
two poles for $\gamma^2$ to take values over a certain range.
\begin{center}
\section{Summary and Discussion}
\end{center}
We have extracted some physical information of sinh(sine)-Gordon model
by using the new R-R sheme. We arrive at an important conclusion which
is substantially different from Ingermanson's that the $D=3$ sinh(sine)
-Gordon model is non-trivial so long as the regularization constant
$\mu^2_s$ is chosen to be $m^2$. This should not be surprising because
for $D=3$, the Coulomb gas model can also be transformed to be a
sine-Gordon model and there should exist a nontrivial quamtum theory for
the
former. Our conclusions agree exactly with those of Ingermanson
  for $D=1,2$ but disagree for $D=3$. Furthermore, our analysis shows
  that for $D=1,2$, the
  RCC has poles for $\gamma^2<0$ and
  the sinh-Gordon
  model has a critical point $\gamma^2_c$ while for $D=3$,
  the RCC has poles for both $\gamma^2>0$ and $\gamma^2<0$.
The existence of the poles of the RCC provides some new mass scales
 as in the
$\lambda\phi^4$ model\cite{s11}. Unfortunately  we  can not still
obtain another critical point $\beta^2_c=4\pi$ which is almost as
important
as $\beta^2_c=8\pi$ in the $D=1$ sine-Gordon model\cite{s7}. This
is perhaps an intrinsic disability of the GEP method.\\
\indent The poles in RCC reflect the intrinsic properties
of the model.
They are neither the mass of solitons nor quite the same
as the so-called "Landau pole $\mu_L$" like that
in QED discussed in previous literatures. In the past, the Landau
pole $\mu_L$ emerges as an singularity or obstruction on the way
of running of cutoff $\Lambda\rightarrow\infty$, or some
arbitrary mass scale $\mu$ ( which stems from some regularization
procedure, e.g. the dimensional regularization) approaching
to infinity. Of course, there is some similarity between
Landau pole and the largest mass scale in our treatment. For
example, in ref.\cite{s11} it is found that there are
three mass scales characterizing the $\lambda\phi^4$ model,
among them, the largest one, say $\mu_c$, can only be found
by non-perturbative method (like GEP) and evolves into
the largest energy scale in the standard model
of particle physics where the $\phi$-field is coupled
to gauge fields. At $\mu_c$, the system undergoes a phase
transition in vacuum (from symmetry broken phase to
symmetric one). We guess that similar phase transition
would occur also in the models considered in this paper.\\
\indent As in the present R-R scheme, there is no explicit
divergence (which is substituted by some constants $C,\mu_s$),
no counterterm, no bare parameter and no arbitrary running
mass scale (all $\mu_i$ in our treatment are fixed and all running
parametres are physical ones) as well. There is no
obtruction in the running of cutoff $\Lambda\rightarrow\infty$
and no bare parametre, say $\gamma_0$ either, so there is
no contradiction enforcing $\gamma_0\rightarrow 0$. Hence
we claim that there is no "triviality" in $D=3$ sinh(sine)-Gordon
model as that in $\lambda\phi^4$ model\cite{s11}. A useful model
should be non-trivial. On the other hand,
very probably it has some
singularities e.g. some poles of RCC, showing the boundary
of its applicability. To know the physics at the singularitis
is beyond the ability of the QFT under consideration. \\
\indent As discussed in ref.\cite{s9}, the QFT is not
well-defined
by the Lagrangian solely. In GEP scheme, a model is defined
by the effective Hamiltonian
\beq
{\cal H}_{eff}=\frac{1}{2}\dot{\Phi}^2+\frac{1}{2}(\nabla\Phi)^2
+V_G(\Phi)
\eeq
with $V_G$ containing some arbitrary constants ($C, \mu_i$). The
constant are the necessary compliments to the original$ {\cal L}$
before the model can be well-defined. They are nothing but the
values of mass scales and coupling constants. In some sense,
the renormalization in QFT is just like to reconfirm the plane
ticket before one's departure from the airport. We must
keep the same symbol of parametres, say $m$, through out
the whole calculation.\\
\indent Once these constants are fixed, the model is well
defined and has some prediction power. The calculation of eq.(74)
at tree level already includes the quantum corrections. We
can consider any momentum dependent vertices after
the first two terms besides $V_G$ in eq.(74) are taken
into account. Everything is unambiguous and
is well-controled. The reason why
an original "non-renormalizable" model becomes renormalizable
in GEP scheme could be
understood by an example in quantum mechanics .
In the Hamiltonian of hydrogen-like atom, if besides
$H_0=\frac{1}{2\mu^2}{\bf p}^2-\frac{Ze^2}{4\pi r}
$, we add a small perturbation term,
$H^{\prime}=Ae^{-bp^2}=A\sum_{k=0}^{\infty}
b^kp^{2k}$, then the energy correction in
eigenstate $\mid nlm>$ remains finite and fixed
to be $\Delta E=<nlm\mid H^{\prime}\mid nlm>$ whereas
the contribution of individual term in $H^{\prime}$,
$<nlm\mid p^{2k}\mid nlm>, (k\ge 3)$, would {\it diverge} !
Once again, this example reminds us of the implication of
divergence, which is by no means a very large number.
Rather, it is essentially a warning, showing that there
might some lack of knowledge or some unsuitability
in our treatment.
For the moment, we can
not claim that what we find is the only finite
solution of the model which was believed as non-renormalizable.
But we think an outcome from GEP manipulation could be
meaningful since the experince in physics often tell
us that the nature does not reject the simpler possibility.\\
\indent In the case of $\gamma^2=-\beta^2<0$, i.e. in the sine-Gordon
model, the original $V(\phi)\sim {\rm cos}\beta\phi$ has the
discrete translational symmetry:$\phi\rightarrow\phi+
\frac{2\pi n}{\beta}$. At first sight, the ansatz of the Gaussian
wave functional Eq.(10) would break this symmetry.
First, in general one
can not expect that the ground state has the same symmetry as
the Hamiltonian \cite{s23}. Note
 that, however, what appears in eq.(10) is the difference
$(\phi_x-\Phi_x)$ not $\phi_x$ itself. Then the contributions
of the fluctuations in different configuration of $\phi$ with
$n\not=0$ are taken into account conceptually for a fixed
$\Phi_x$ in the path integral. Yet, the contributions
for $n\not=0$ is strongly suppressed. In ref.\cite{s24}(see
also \cite{s3}), the soliton linking neighbouring $\Phi$ sectors
in quantized sine-Gordon model is considered in the $D=1$ case
with the GEP as shown here by eqs.(30),(47) and (53)
\beq
V_G(\Phi)=\frac{m^2_R}{\beta^2}(\frac{\beta^2}{8\pi}-1)({\rm cos}
\beta\Phi)^{8\pi/(8\pi-\beta^2)}
\eeq
which still preserves the symmetry.
For the $D=2$ (or 3) case, through we can not write down
an explicit GEP like eq.(75) due to the complicate gap
equation (63) (or (72)), we are still able to see that
the GEP preserves the periodic symmetry,i.e.
\beq
V_G(\Phi)=V_G(\Phi+\frac{2n\pi}{\beta})
\eeq
\indent In summary, the GEP approach combining with
the new R-R method does provide a nice calculational
scheme for non-perturbative QFT.\\
\underline{\bf Acknowledgement} This work was supported by the Funds
for Young Teachers of Education Commitee of Shanghai and the National
Science Foundation of China under Grant. No. 19805004.

\end{document}